\documentclass[prl,twocolumn,nofootinbib,aps,showpacs]{revtex4-1}

\usepackage{amsmath}
\usepackage{amssymb}
\usepackage{graphicx}
\usepackage{color}
\usepackage{bm}
\usepackage{times}
\usepackage{hyperref}
\hypersetup{
  colorlinks=true,
  citecolor=blue,
  linkcolor=blue,
  urlcolor=blue}

\usepackage{soul}

\everymath{\displaystyle}

\usepackage{epsfig}
\usepackage{dcolumn}
\usepackage{float}

\newcommand{\dcp}{\delta_\text{CP}}
\newcommand{\nova}{NO$\nu$A\ }

\begin{document}

\renewcommand{\arraystretch}{1.3}

\begin{titlepage}

\title{Is nonstandard interaction a solution to the
three neutrino tensions?}

\author{Shinya Fukasawa}
\email[Email Address: ]{fukasawa-shinya@ed.tmu.ac.jp}
\affiliation{Department of Physics, Tokyo Metropolitan University, Hachioji, Tokyo 192-0397, Japan}

\author{Monojit Ghosh}
\email[Email Address: ]{monojit@tmu.ac.jp}
\affiliation{Department of Physics, Tokyo Metropolitan University, Hachioji, Tokyo 192-0397, Japan}

\author{Osamu Yasuda}
\email[Email Address: ]{yasuda@phys.se.tmu.ac.jp}
\affiliation{Department of Physics, Tokyo Metropolitan University, Hachioji, Tokyo 192-0397, Japan}

\begin{abstract}
In this work we present a scenario in which
a nonstandard interaction in neutrino
propagation can explain the three major tensions in the neutrino oscillation data at present. These tensions are: (i) a non-zero best-fit value of the non-standard oscillation parameters in
the the global analysis of the solar and KamLAND data which rules out the standard oscillation scenario at $90\%$ C.L, (ii) the measurement of the non-maximal value of $\theta_{23}$ by \nova which
excludes the maximal mixing at $2.5 \sigma$ C.L. and (iii) a
discrepancy in the $\theta_{13}$ measurement by T2K which has a
tension with the reactor best-fit value of $\sin^2\theta_{13}=0.021$ at $90\%$ C.L.
Our results show that all these three above mentioned anomalies can be explained if one assumes the existence of the non-standard interactions in neutrino propagation with $\theta_{23}=45^\circ$ and 
$\sin^2\theta_{13}=0.021$ in the case of normal hierarchy.
In our scenario the phase of $\epsilon_{e\tau}$ is zero
and the most favorable value of
the Dirac CP phase is approximately $255^\circ$.
\end{abstract}

\maketitle

\end{titlepage}

Neutrino oscillation experiments have been successful in determination
of the three mixing angles ($\theta_{12}$, $\theta_{13}$ and $\theta_{23}$) and the two mass squared differences ($\Delta m_{21}^2$ and $\Delta m_{31}^2$).  What
remains to be studied is the mass hierarchy of neutrinos (either normal (NH): $\Delta m_{31}^2 >0 $ or inverted (IH):  $\Delta m_{31}^2 < 0$), the precise value of the mixing angle $\theta_{23}$ and the
Dirac CP phase $\dcp$. There are many future experiments planned to determine
these unknown quantities.  In the mean time, a few tensions in neutrino
experiments have been reported recently. They are: (i) 
the tension between the mass squared differences by the
solar and KamLAND data\,\cite{Gonzalez-Garcia:2013usa} which gives a non-zero best-fit value of the non standard interaction parameters $\epsilon_D$ and $\epsilon_N$. This rules out the standard oscillation scenario at
$90\%$ C.L, (ii) the tension between the T2K and \nova experiments regarding the measurement of the mixing angle $\theta_{23}$ \cite{t2k,nova} and (iii) the tension in the measurement of the mixing angle
$\theta_{13}$ by the reactor and T2K experiments\,\cite{Abe:2015awa,An:2015rpe}.\footnote{
The latest measurement by Daya Bay\,\cite{An:2015rpe}
gives $\sin^2\theta_{13}=0.021$ and this lies within 90\%CL of the T2K allowed region
(See Fig. 31 of Ref.\,\cite{Abe:2015awa}).
Although this may not be called a tension
at present, if this trend persists as the
statistics increases, the discrepancy between
the mixing angles
$\theta_{13}$ by the reactor and T2K
experiments should be taken
seriously in future.}
In Table. \ref{tab1} we summarize the recent data of T2K and \nova. 
According to Ref. \cite{Abe:2015awa}, T2K has observed a total of 28 events in the appearance channel and 120 events in the disappearance channel with a
total POT (protons on target) of $6.6 \times 10^{20}$ in the neutrino mode \footnote{The recent update of the T2K data can be found in Ref. \cite{t2k_latest}.
As the details of the fit are not available yet we take
the latest published results for our analysis.}. On the other hand \nova has seen 33 events in the appearance mode and 78 events in the disappearance mode with an exposure of
$6.05 \times 10^{20}$ POT in the neutrino mode \cite{nova}. From Table \ref{tab1}, the tension between the T2K and \nova data are clearly visible. Regarding $\theta_{13}$, 
T2K does its own fit and the best-fit value is
much higher than reactor best fit which is $\sin^2\theta_{13}=0.021$. For $\theta_{23}$, T2K data predicts maximal mixing. On the other hand \nova uses the reactor best-fit value for fitting and it 
excludes maximal mixing for $\theta_{23}$ at $2.5 \sigma$ C.L. and gives a best-fit of $\sin^2\theta_{23}=0.4$.     

In this Letter we look for a scenario which
solves all these three tensions
by introducing a flavor-dependent neutral
current Non-Standard Interaction (NSI) in neutrino
propagation\,\cite{Wolfenstein:1977ue,Guzzo:1991hi,Roulet:1991sm,Ohlsson:2012kf,Miranda:2015dra}\footnote{For recent studies of NSI in long-baseline experiments see \cite{nsi_recent}. }.
The purpose of our work is not to exhaust the whole
parameter space but to show the existence
of a new solution.

The NSI which we discuss here is described by the effective
Lagrangian
\begin{eqnarray}
{\cal L}_{\mbox{\rm\scriptsize eff}}^{\mbox{\tiny{\rm NSI}}} 
=-2\sqrt{2}\, \epsilon_{\alpha\beta}^{fP} G_F
\left(\overline{\nu}_{\alpha L} \gamma_\mu \nu_{\beta L}\right)\,
\left(\overline{f}_P \gamma^\mu f_P\right),
\label{NSIop}
\end{eqnarray}
where $f_P$ stands for fermions with chirality $P$ and
$\epsilon_{\alpha\beta}^{fP}$ is a dimensionless constant
which is normalized by the Fermi coupling constant $G_F$.
In the presence of this NSI, the neutrino evolution is
governed by the Dirac equation:

\begin{eqnarray}
&{\ }&\hspace{-1mm}
i {d \over dx} \left( \begin{array}{c} \nu_e(x) \\ \nu_{\mu}(x) \\ 
\nu_{\tau}(x)
\end{array} \right)
\nonumber\\
&{\ }&\hspace{-6mm}
 = 
\left[  U {\rm diag} \left(0, \frac{\Delta m_{21}^2}{2E},
\frac{\Delta m_{31}^2}{2E}
\right)  U^{-1}
+{\cal A}\right]
\left( \begin{array}{c}
\nu_e(x) \\ \nu_{\mu}(x) \\ \nu_{\tau}(x)
\end{array} \right)\,,
\label{eqn:sch}
\end{eqnarray}
where
\begin{eqnarray}
&{\ }&\hspace{-16mm}
{\cal A} \equiv
A \left(
\begin{array}{ccc}
1+ \epsilon_{ee} & \epsilon_{e\mu} & \epsilon_{e\tau}\\
\epsilon_{\mu e} & \epsilon_{\mu\mu} & \epsilon_{\mu\tau}\\
\epsilon_{\tau e} & \epsilon_{\tau\mu} & \epsilon_{\tau\tau}
\end{array}
\right),
\label{matter-np}
\end{eqnarray}
\fussy
$A\equiv \sqrt{2} G_F N_e$,
$U$ is the leptonic mixing matrix,
$\Delta m_{jk}^2\equiv m_j^2-m_k^2$,
$\epsilon_{\alpha\beta}$ is defined by
\begin{eqnarray}
&{\ }&\hspace{-36mm}
\epsilon_{\alpha\beta}\equiv\sum_{f=e,u,d}\frac{N_f}{N_e}\epsilon_{\alpha\beta}^{f}.
\label{epsilon1}
\end{eqnarray}
We defined the new NSI parameters as
$\epsilon_{\alpha\beta}^{f}\equiv\epsilon_{\alpha\beta}^{fL}+\epsilon_{\alpha\beta}^{fR}$
since the matter effect is sensitive only to the coherent scattering
and only to the vector part in the interaction,
and $N_f~(f=e, u, d)$ stands for the number densities of fermions $f$.
\begin{table}
\begin{center}
\begin{tabular}{cccc}
\hline
Expt   & $\sin^2\theta_{13}$ NH (IH)  & $\sin^2\theta_{23}$ NH (IH)  & $\dcp$ NH (IH)   \\          
\hline
T2K  &    0.0422 (0.0491)   & 0.524 (0.523)    &  1.91 (1.01)     \\
\nova      &   0.021   &  0.040    &  1.49 $\pi$  \\
\hline
\end{tabular}
\end{center}
\caption{Recent data of T2K and \nova.}
\label{tab1} 
\end{table}

To discuss the effect of NSI on solar neutrinos,
the $3 \times 3$ Hamiltonian in the Dirac equation
Eq.\,(\ref{eqn:sch}) is
reduced to an effective $2 \times 2$ Hamiltonian given by
\begin{eqnarray}
&{\ }&\hspace{-6mm}
H^{\rm eff}=
\frac{\Delta m^2_{21}}{4E}\left(\begin{array}{cc}
-\cos2\theta_{12} & \sin2\theta_{12}  \\
\sin2\theta_{12} & \cos2\theta_{12}
\end{array}\right) 
\nonumber\\
&{\ }&\hspace{4mm}
+
\left(\begin{array}{cc}
c^2_{13} A & 0 \\
0 & 0
\end{array}\right)  + 
 A\sum_{f=e,u,d} \frac{N_f}{N_e}
\left(\begin{array}{cc}
- \epsilon_{D}^f &  \epsilon_{N}^f \\
 \epsilon_{N}^{f*} &  \epsilon_{D}^f
\end{array}\right),
\end{eqnarray}
where  $\epsilon^f_{D}$ and $\epsilon^f_{N}$ are linear combinations of the standard NSI parameters:
\begin{eqnarray}
&{\ }&\hspace{-6mm}
\epsilon_{D}^f 
=
-\frac{c_{13}^2}{2}\left(\epsilon_{e e}^f-\epsilon_{\mu \mu}^f\right)+\frac{s_{23}^2-s_{13}^2c_{23}^2}{2}\left(\epsilon_{\tau \tau}^f-\epsilon_{\mu \mu}^f\right) 
\nonumber\\
&{\ }&\hspace{4mm}
+c_{13}s_{13}{\rm Re}\left[ e^{i\delta_{\rm CP}}\left(s_{23}\epsilon_{e \mu}^f+c_{23}\epsilon_{e \tau}^f\right) \right]
\nonumber\\
&{\ }&\hspace{2mm}
-\left(1+s_{13}^2\right)c_{23}s_{23}{\rm Re}\left[\epsilon_{\mu \tau}^f\right]
\label{epsilond}\\
&{\ }&\hspace{-6mm}
\epsilon_{N}^f= -c_{13}s_{23}\epsilon_{e\tau}^f
\nonumber\\
&{\ }&\hspace{2mm}
+c_{13}c_{23}\epsilon_{e \mu}^f
+s_{13}c_{23}s_{23}e^{-i\delta_{\rm CP}}
\left(\epsilon_{\tau \tau}^f-\epsilon_{\mu \mu}^f\right) 
\nonumber\\
&{\ }&\hspace{2mm}
+s_{13}e^{-i\delta_{\rm CP}}\left( s_{23}^2\epsilon_{\mu\tau}^f-c_{23}^2\epsilon_{\mu\tau}^{f*} \right)
\,,
\label{epsilonn}
\end{eqnarray}
and $c_{jk}\equiv\cos\theta_{jk}$, $s_{jk}\equiv\sin\theta_{jk}$.
In the analysis of Ref.\,\cite{Gonzalez-Garcia:2013usa},
one particular choice of $f=u$ or $f=d$ was taken at a time
because of the nontrivial composition profile of the Sun,
and it was found that the best fit values are
$(\epsilon_{D}^u,\epsilon_{N}^u)=(-0.22,-0.30)$ ($f=u$) or
$(\epsilon_{D}^d,\epsilon_{N}^d)=(-0.12,-0.16)$ ($f=d$) 
from the solar neutrino and KamLAND data only,
$(\epsilon_{D}^u,\epsilon_{N}^u)=(-0.140,-0.030)$ ($f=u$) or
$(\epsilon_{D}^d,\epsilon_{N}^d)=(-0.145,-0.036)$ ($f=d$) 
from the global analysis of the neutrino oscillation data.
\begin{figure*}
\begin{center}
\begin{tabular}{lr}
\includegraphics[scale=1.0]{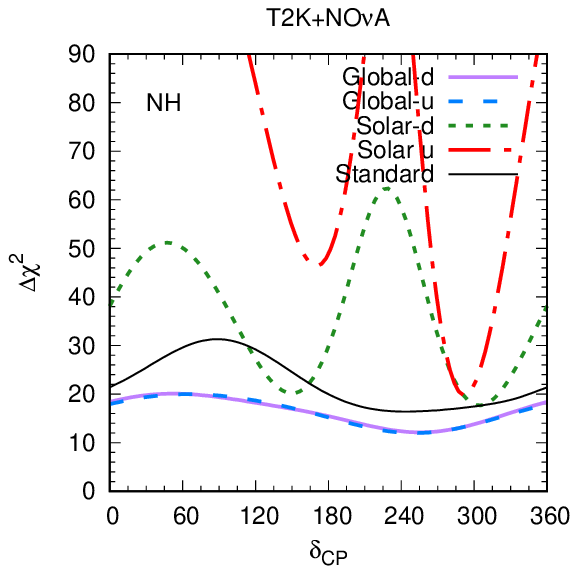}
\hspace{-0.8 in}
\includegraphics[scale=1.0]{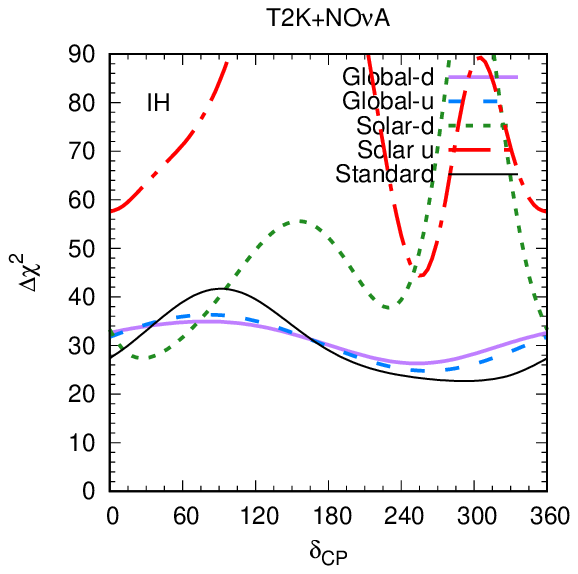}
\end{tabular}
\end{center}
\caption{
The significance of the four best-fit solutions of
the solar+KamLAND data (or the global analysis
including the solar+KamLAND data;
for $f=u$ or $f=d$)
for the combined fit to the
T2K and \nova data for NH (left panel)
and IH (right panel).  The black solid curve
corresponds to the standard case with
$\theta_{23}=45^\circ$ and $\sin^2\theta_{13}=0.021$ without NSI.
}
\label{fig1}
\end{figure*}
%
\begin{figure}
\begin{center}
\begin{tabular}{lr}
\includegraphics[scale=1.0]{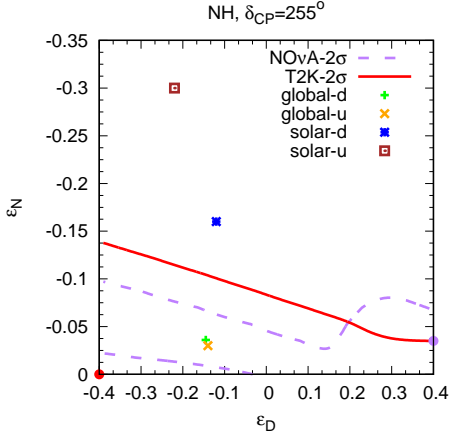}
\end{tabular}
\end{center}
\caption{Allowed region in the $\epsilon_D$ - $\epsilon_N$ plane. The best-fit points for \nova and T2K are represented by the purple and red dot respectively.}
\label{fig2}
\end{figure}

In this work we look for a scenario with NSI
which gives a good fit to the solar and KamLAND data,
the \nova data and the T2K data.
For our analysis we use the GLoBES \cite{Huber:2004ka} and MonteCUBES \cite{Blennow:2009pk} softwares.
For our fit we will assume that
the mixing angle $\theta_{23}$ in vacuum is
maximal, i.e., $\theta_{23}=45^\circ$\,\footnote{
A similar attempt was made in Ref.\,\cite{Yasuda:2010aa}
to use NSI to reconcile
the different values of $\theta_{23}$ for
the neutrino and antineutrino modes.} and
the mixing angle $\theta_{13}$ in vacuum is
given by the reactor data, i.e.,
$\sin^2\theta_{13}=0.021$. 
We will do our analysis in the $(\epsilon_{D}^f,\epsilon_{N}^f)$ plane and for this we need to express $\epsilon_{\alpha\beta}$ as a function of $(\epsilon_{D}^f,\epsilon_{N}^f)$.
So we proceed in the following way.
As can be seen from
the definition of $\epsilon_{\alpha\beta}$, the neutrino oscillation
experiments on the Earth are sensitive only to the sum of
$\epsilon_{\alpha\beta}^{f}$.  However, since
the analysis of solar neutrinos was done either
for $f=u$ or $f=d$ only, we also analyze the
long baseline experiments assuming the
same condition.
Since the number of neutrons and that of electron
is approximately equal in the Earth, if we turn on
NSI for $f=u$ only or $f=d$ only, then from Eq.\,(\ref{epsilon1})
we get
\begin{eqnarray}
&{\ }&\hspace{-36mm}
\epsilon_{\alpha\beta}=3\epsilon_{\alpha\beta}^{f}\,.
\label{epsilon2}
\end{eqnarray}
As we can see from Eqs.\,(\ref{epsilond}) and (\ref{epsilonn}),
the mapping $\epsilon_{\alpha\beta}^{f}$ $\to$ $(\epsilon_{D}^f, \epsilon_{N}^f)$
is not one to one, and in general it is difficult to obtain
the possible region for
the $\epsilon_{\alpha\beta}$ parameters analytically.
Here, instead of exhausting all the possible
regions for $\epsilon_{\alpha\beta}$, we postulate the
following:
\begin{eqnarray}
&{\ }&\hspace{-6mm}
\epsilon_{D}^f 
=
-\frac{c_{13}^2}{2}\left(\epsilon_{e e}^f-\epsilon_{\mu \mu}^f\right)+\frac{s_{23}^2-s_{13}^2c_{23}^2}{2}\left(\epsilon_{\tau \tau}^f-\epsilon_{\mu \mu}^f\right) 
\label{epsilond1}\\
&{\ }&\hspace{-4mm}
0=c_{13}s_{13}{\rm Re}\left[ e^{i\delta_{\rm CP}}\left(s_{23}\epsilon_{e \mu}^f+c_{23}\epsilon_{e \tau}^f\right) \right]
\nonumber\\
&{\ }&\hspace{2mm}
-\left(1+s_{13}^2\right)c_{23}s_{23}{\rm Re}\left[\epsilon_{\mu \tau}^f\right]
\label{epsilond2}\\
&{\ }&\hspace{-6mm}
\epsilon_{N}^f= -c_{13}s_{23}\epsilon_{e\tau}^f
\label{epsilonn1}\\
&{\ }&\hspace{-4mm}
0=c_{13}c_{23}\epsilon_{e \mu}^f
+s_{13}c_{23}s_{23}e^{-i\delta_{\rm CP}}
\left(\epsilon_{\tau \tau}^f-\epsilon_{\mu \mu}^f\right) 
\nonumber\\
&{\ }&\hspace{2mm}
+s_{13}e^{-i\delta_{\rm CP}}\left( s_{23}^2\epsilon_{\mu\tau}^f-c_{23}^2\epsilon_{\mu\tau}^{f*} \right)
\,,
\label{epsilonn2}
\end{eqnarray}
Furthermore, for simplicity, we postulate
Im\,($s_{23}^2\epsilon_{\mu\tau}^f-c_{23}^2\epsilon_{\mu\tau}^{f*}$) = 0,
which implies that $\epsilon_{\mu\tau}^f$
is a real parameter in the case
of $\theta_{23}=45^\circ$,
and following the bound from
the high energy atmospheric neutrino data,
we take\,\cite{Friedland:2004ah,Friedland:2005vy}
\begin{eqnarray}
&{\ }&\hspace{-21mm}
\epsilon_{\tau\tau}=\frac{|\epsilon_{e\tau}|^2}{1+\epsilon_{ee}}\,.
\label{eq:ansatz_a}
\end{eqnarray}
Another constraint comes from the
atmospheric neutrino data, and
the following must be satisfied:\,\cite{Fukasawa:2015jaa}
\begin{eqnarray}
&{\ }&\hspace{-11mm}
\left|\frac{\epsilon_{e\tau}}{1+\epsilon_{ee}}\right|
\lesssim 0.8\quad\mbox{\rm at}~2.5\sigma\mbox{\rm CL}\,.
\label{tanb}
\end{eqnarray}
Finally, we put $\epsilon_{\mu\mu}=0$ because
we can always redefine
$\epsilon_{ee}-\epsilon_{\mu\mu}\to\epsilon_{ee}$ and
$\epsilon_{\tau\tau}-\epsilon_{\mu\mu}\to\epsilon_{\tau\tau}$.
From these assumptions and
Eqs.\,(\ref{epsilon2}), (\ref{epsilond1}), (\ref{epsilond2}), 
(\ref{epsilonn1}), (\ref{epsilonn2}), (\ref{eq:ansatz_a}) and (\ref{tanb}),
after putting $\theta_{23}=45^\circ$,
we get the
following expressions:
\begin{eqnarray}
&{\ }&\hspace{-1mm}
\epsilon_{e\tau}=-\frac{3\sqrt{2}}{c_{13}}\epsilon_{N}^f
\label{epsilonet}\\
&{\ }&\hspace{-1mm}
\epsilon_{ee}=-\frac{3}{c_{13}^2}\epsilon_{D}^f-\frac{1}{2}
+\left\{
\left(\frac{3}{c_{13}^2}\epsilon_{D}^f-\frac{1}{2}
\right)^2+\frac{1}{c_{13}^2}\left|3\epsilon_{N}^f\right|^2
\right\}^{1/2}
\nonumber\\
\label{epsilonee}
\\
&{\ }&\hspace{-1mm}
\epsilon_{e\mu}=-\frac{s_{13}}{\sqrt{2}\,c_{13}}e^{-i\dcp}
\frac{|\epsilon_{e\tau}|^2}{1+\epsilon_{ee}}
\label{epsilonem}
\\
&{\ }&\hspace{-1mm}
\epsilon_{\mu\tau}=\frac{\sqrt{2}}{1+s_{13}^2}
c_{13}s_{13}\mbox{\rm Re}\left[
e^{i\dcp}\left(\epsilon_{e\mu}+\epsilon_{e\tau}
\right)
\right]
\label{epsilonmt}
\end{eqnarray}
Note that in all the best fit solutions from
the solar+KamLAND analysis,
both $\epsilon_{D}^f$ and $\epsilon_{N}^f$ have
a phase $(-1)$, so in the present ansatz
we have arg($\epsilon_{e\tau}$)=0 from
Eq.\,(\ref{epsilonet}), and 
arg($\epsilon_{e\mu}$) = $\pi-\dcp$ from
Eq.\,(\ref{epsilonem}).

Although our ansatz gives us
only a special solution to
Eqs.\,(\ref{epsilond}) and (\ref{epsilonn}),
it covers some of the whole solution space
in the following way.
We have verified that 
the appearance
probability $P(\nu_\mu\to\nu_e)$ for \nova
and T2K is not very
sensitive to the small parameters
$\epsilon_{e\mu}$ and $\epsilon_{\mu\tau}$.
So even if we vary the $\epsilon_{\alpha\beta}$
in general, the behavior of the fit
is not expected to be very much different
from what is obtained by our ansatz.

Thus we obtained the values for the $\epsilon_{\alpha\beta}$ parameters which depend on one free parameter $\dcp$. As we mentioned earlier there are four best-fit points ($\epsilon^{\rm sol}$) for the solar data. 
For our fit we calculated the $\chi^2$ at each solar best-fit point for T2K and \nova assuming $\theta_{23}^{\rm fit}=45^\circ$ and $\sin^2\theta_{13}^{\rm fit} = 0.021$ using the following formula:
\begin{eqnarray}
&{\ }&\hspace{-5mm}
\chi^2 (\dcp) \equiv
\sum_j \frac{1}{N_j^{\rm data}}
\left[N_j^{\rm th}(\epsilon^{\rm sol}, \theta_{23}^{\rm fit},\theta_{13}^{\rm fit})
-N_j^{\rm data}\right]^2,
\label{hierarchy-chi2}
\end{eqnarray}
where for `data' we take the numbers as given in Table. \ref{tab1}.
In the analysis, we introduce
the prior for $\epsilon_{e\mu}$ and $\epsilon_{\mu\tau}$:
\begin{eqnarray}
&{\ }&\hspace{-5mm}
\chi_\text{prior}^2
 = 2.7\left(\frac{|\epsilon_{e\mu}|}{0.15}\right)^2
   + 2.7\left(\frac{|\epsilon_{\mu\tau}|}{0.15}\right)^2\,,
\label{chiprior}
\end{eqnarray}
where the bound for each parameter at 90\%CL
was taken from Ref.\,\cite{Biggio:2009nt}.
In the combined analysis, we evaluate
the total $\chi^2$ given by
\begin{eqnarray}
&{\ }&\hspace{-5mm}
\chi^2
 = \chi_\text{nova}^2 + \chi_\text{T2K}^2 + \chi_\text{solar+KL}^2
 + \chi_\text{prior}^2
\label{chitotal}
\end{eqnarray}
for all the values of $\dcp$ and plot in Fig. \ref{fig1} for both the hierarchies.
In Eq.\.(\ref{chitotal}) we approximated $\chi_\text{solar+KL}^2$
as $\chi_\text{solar+KL}^2\simeq \chi^2(\epsilon_D)+\chi^2(\epsilon_N)$,
where $\chi^2(\epsilon_D)$ and $\chi^2(\epsilon_N)$
are $\chi^2$ obtained from the solar+KamLAND data
in Ref.\,\cite{Gonzalez-Garcia:2013usa}.
Obviously for the solar+KamLAND best-fit points,
$\chi_\text{solar+KL}^2=0$, while for the global
best-fit points, $\chi_\text{solar+KL}^2=0.1$.
The latter was estimated from the Figure.2 of
Ref.\,\cite{Gonzalez-Garcia:2013usa}.
To estimate the goodness of fit, we compare our $\chi^2$ with $\chi^\text{2 std}$, i.e., the standard case.
By $\chi^\text{2 std}$ we mean the value of $\chi^2$ at ($\theta_{23}^{\rm fit},\theta_{13}^{\rm fit}$) without NSI. 
Here the $\chi^\text{2 std}$ for solar+KamLAND
is 3.8 (4.4) for $f=u$ ($f=d$), which is estimated
by the approximation mentioned earlier.
In our analysis we take the value $\chi^\text{2 std}_\text{solar+KL}=3.8$
for conservative estimation.
On the other hand, the $\chi^\text{2 std}$ of
T2K and \nova 
depend on
$\dcp$.  For the standard case,
therefore, we have
$\chi_\text{nova}^\text{2 std} + \chi_\text{T2K}^\text{2 std} + 3.8$.

From Fig. \ref{fig1} we see the following. 
For NH the two curves (solid-purple and dashed-blue) which correspond to the best-fit points of the
global analysis of the solar data lie below the standard curve (solid-black) for 
all the values of $\dcp$.
This implies that a nonstandard interaction at the solar best-fit point gives a better fit as compared to the standard case.
Thus we found a new solution (the best-fit point of the global analysis of the solar+kamLAND data) with NSI which solves all the three neutrino tensions.
 Whereas in IH, a scenario with NSI in any region of 
$\dcp$ does not give $\chi^2$ which is smaller than the minimum $\chi^2$ in the standard case. From the plot we also see that
in the case of NH, $\dcp\simeq 255^\circ$
is the most preferred value of $\dcp$ which gives the best fit with NSI.
In Table \ref{tab2} we give the values for
$\epsilon_{\alpha\beta}$ corresponding to the global-$u$ best-fit point of the solar data at $\dcp=255^\circ$.
\begin{table}
\begin{center}
\begin{tabular}{ccccc}
\hline
 $\epsilon_{ee}$  & $\epsilon_{e\tau}$  & $\epsilon_{\tau \tau}$ & $|\epsilon_{e \mu}|$ & $\epsilon_{\mu \tau}$\\          
\hline
  0.84885   & 0.12863     &  0.008950 & 0.00092689 & -0.0067963    \\
\hline
\end{tabular}
\end{center}
\caption{Values of $\epsilon_{\alpha \beta}$ corresponding to global-$u$ best fit point of the solar data ($\epsilon_D=-0.14$, $\epsilon_N=-0.03$) at $\dcp=255^\circ$.}
\label{tab2} 
\end{table}
For our information, in Fig. \ref{fig2} we give the allowed region in the
($\epsilon_D$, $\epsilon_N$) plane for \nova and T2K
at $\dcp= 255^\circ$ in the case of NH.
As mentioned earlier,
since $\epsilon_{N}^f$ ($\epsilon_{e\mu}$)
has a phase $(-1)$ ($0$)
in all the best-fit solutions from the solar+KamLAND analysis,
we performed our analysis only for $\epsilon_{N}^f<0$.
For the solar+KamLAND we give just the best-fit points.
From these plots we identify the allowed region which is consistent with all the three anomalies under discussion. 
For NH we see that the global best-fit of the solar data is consistent with the \nova and T2K data within $2 \sigma$ confidence regions.

In summary we found a scenario
which explains the tension of the
mass squared differences of the
solar and KamLAND data, the one of mixing
angles $\theta_{23}$ of the T2K and \nova data,
and the discrepancy of $\theta_{13}$
of the reactor and T2K data.
In our analysis we found that
the goodness of fit for the NSI scenario is better for all the values of $\dcp$ in NH as compared to the standard case and that
$\dcp\simeq 255^\circ$ give a bet-fit among others.
For IH the NSI does not give a better fit.
In this scenario,
the three tensions give a constraint
on the phase of $\epsilon_{e\tau}$ as zero
and the most favorable value of
the Dirac CP phase is $\simeq 255^\circ$.
To be conclusive, we need more statistics
of the T2K and \nova experiments.
If the best fit values for $\theta_{23}$
at both the T2K and \nova experiments
or the best fit values for $\theta_{13}$
of the reactor and T2K data
remain the same as the statistics increases,
then the present scenario with NSI will
give a better fit to the data.
It should be pointed out that
the solar neutrino
observation at the Hyperkamokande
experiment is expected to
test the tension between
the solar and KamLAND data by the day night
effect\,\cite{kajita},
and also that the atmospheric neutrino
observation at the Hyperkamokande
experiment is expected to test this NSI scenario
by the matter effect in the multi-GeV energy
range.\,\cite{Fukasawa:2016nwn}

Towards the completion of this work,
we became aware of Ref.\,\cite{Liao:2016bgf},
which discussed part of the ideas in our paper
from a different point of view.

This research was partly supported by a Grant-in-Aid for Scientific
Research of the Ministry of Education, Science and Culture, under
Grants No. 25105009, No. 15K05058, No. 25105001 and No. 15K21734.

\bibliographystyle{apsrev}

\begin{thebibliography}{99}

\bibitem{Gonzalez-Garcia:2013usa} 
  M.~C.~Gonzalez-Garcia and M.~Maltoni,
  JHEP {\bf 1309}, 152 (2013)
  [arXiv:1307.3092].

\bibitem{t2k} 
\fussy
L. Magaletti, talk at NOW2016,
Otranto, Italy, 4 -- 11 September, 2016.

\bibitem{nova} 
\fussy
P. Vahle, talk at Neutrino 2016,
4-9 July, London.

\bibitem{Abe:2015awa} 
  K.~Abe {\it et al.} [T2K Collaboration],
  Phys.\ Rev.\ D {\bf 91}, no. 7, 072010 (2015)
  [arXiv:1502.01550 [hep-ex]].

\bibitem{An:2015rpe} 
  F.~P.~An {\it et al.} [Daya Bay Collaboration],
  Phys.\ Rev.\ Lett.\  {\bf 115}, no. 11, 111802 (2015)
  [arXiv:1505.03456 [hep-ex]].

  \bibitem{t2k_latest} 
\fussy
H. A. Tanaka, talk at Neutrino 2016,
4-9 July, London.
  
\bibitem{Wolfenstein:1977ue}
  L.~Wolfenstein,
  Phys.\ Rev.\  D {\bf 17}, 2369 (1978).

\bibitem{Guzzo:1991hi}
  M.~M.~Guzzo, A.~Masiero and S.~T.~Petcov,
  Phys.\ Lett.\  B {\bf 260} (1991) 154.

\bibitem{Roulet:1991sm} 
  E.~Roulet,
  Phys.\ Rev.\ D {\bf 44}, R935 (1991). 
  
\bibitem{Ohlsson:2012kf}
  T.~Ohlsson,
  Rept.\ Prog.\ Phys.\  {\bf 76} (2013) 044201
  [arXiv:1209.2710 [hep-ph]].

\bibitem{Miranda:2015dra}
  O.~G.~Miranda and H.~Nunokawa,
  New J.\ Phys.\  {\bf 17} (2015) no.9,  095002
  [arXiv:1505.06254 [hep-ph]].
  
\bibitem{nsi_recent} 
  M.~Blennow, S.~Choubey, T.~Ohlsson, D.~Pramanik and S.~K.~Raut,
  JHEP {\bf 1608}, 090 (2016)
  [arXiv:1606.08851 [hep-ph]],
    D.~V.~Forero and P.~Huber,
  Phys.\ Rev.\ Lett.\  {\bf 117}, 031801 (2016)
  [arXiv:1601.03736 [hep-ph]],
  P.~Coloma and T.~Schwetz,
  Phys.\ Rev.\ D {\bf 94}, 055005 (2016)
  [arXiv:1604.05772 [hep-ph]],
  M.~Masud, A.~Chatterjee and P.~Mehta,
  J.\ Phys.\ G {\bf 43}, no. 9, 095005 (2016)
  [arXiv:1510.08261 [hep-ph]],
  S.~K.~Agarwalla, S.~S.~Chatterjee and A.~Palazzo,
  arXiv:1607.01745 [hep-ph],
  A.~de Gouvêa and K.~J.~Kelly,
  Nucl.\ Phys.\ B {\bf 908}, 318 (2016)
  [arXiv:1511.05562 [hep-ph]].

\bibitem{Huber:2004ka} 
  P.~Huber, M.~Lindner and W.~Winter,
  Comput.\ Phys.\ Commun.\  {\bf 167}, 195 (2005)
  [hep-ph/0407333].
  
\bibitem{Blennow:2009pk} 
  M.~Blennow and E.~Fernandez-Martinez,
  Comput.\ Phys.\ Commun.\  {\bf 181}, 227 (2010)
  [arXiv:0903.3985 [hep-ph]].

\bibitem{Yasuda:2010aa} 
  O.~Yasuda,
  AIP Conf.\ Proc.\  {\bf 1382}, 103 (2011)
  [arXiv:1012.3478 [hep-ph]].
  
\bibitem{Friedland:2004ah}
  A.~Friedland, C.~Lunardini and M.~Maltoni,
  Phys.\ Rev.\  D {\bf 70}, 111301 (2004)
  [arXiv:hep-ph/0408264].

\bibitem{Friedland:2005vy}
  A.~Friedland and C.~Lunardini,
  Phys.\ Rev.\  D {\bf 72} (2005) 053009
  [arXiv:hep-ph/0506143].
  
\bibitem{Fukasawa:2015jaa} 
  S.~Fukasawa and O.~Yasuda,
  Adv.\ High Energy Phys.\  {\bf 2015}, 820941 (2015)
  [arXiv:1503.08056 [hep-ph]].

\bibitem{Biggio:2009nt}
  C.~Biggio, M.~Blennow and E.~Fernandez-Martinez,
  JHEP {\bf 0908}, 090 (2009)
  [arXiv:0907.0097 [hep-ph]].

\bibitem{kajita} 
\fussy
T. Kajita, talk at NOW2016,
Otranto, Italy, 4 -- 11 September, 2016.

\bibitem{Fukasawa:2016nwn} 
  S.~Fukasawa and O.~Yasuda,
  arXiv:1608.05897 [hep-ph].

\bibitem{Liao:2016bgf} 
  J.~Liao, D.~Marfatia and K.~Whisnant,
  arXiv:1609.01786 [hep-ph].

\end{thebibliography}

\end{document}